\documentstyle[cjaa209,twoside,epsf]{article}
\input{psfig.sty}

\begin{document}

\title{Low Frequency Radio Observations of \\ GRS1915+105 with GMRT}
\author{C. H. Ishwara-Chandra$^{1\star}$, A. Pramesh Rao$^1$, Mamta Pandey$^2$, 
R. K. Manchanda$^3$, Philippe Durouchoux$^4$}
\inst{$^1$ National Center for Radio Astrophysics, TIFR, P. B. No. 3,
Ganeshkhind, Pune - 7, India  \\
$^2$ Department of Physics, University of Mumbai, 
Mumbai - 400 078, India \\
$^3$ Tata Institute of Fundamental Research, Homi Bhaba Road,
Mumbai - 400 005, India \\
$^4$ CEN Saclay, DSM, DAPNIA, Service d'Astrophysique, 
91191 Gif sur Yvette, France
\email{ishwar@ncra.tifr.res.in}
}

\date{}

\abstract{ 

We present the first detailed low frequency radio measurements of the
galactic microquasar GRS1915+105 with GMRT. Simultaneous observations
were carried out at 610 and 244 MHz. Our data does not show any
signature of spectral turn over even at low radio frequency of 244 MHz.
We propose that while the radio emission at high radio frequencies could
predominantly come from compact jets, the emission at lower frequency
originates in the lobes at the end of the jet which acts like a reservoir
of low energy electrons.

}


\section{Introduction}

The Galactic X-ray transient source GRS1915+105 was discovered in
1992 by the {\tt GRANAT} satellite (Castro-Tirado et al. 1992). It has
now been established as a black hole binary with the black hole mass
of 14$\pm 4 M_\odot$ and the companion K-M III star of mass 1.2$\pm
0.2 M_\odot$ (Greiner et al. 2001). The source was shown to possess
relativistic outflows during its radio outbursts (Mirabel \& Rodr\'iguez
1994). Extensive monitoring of GRS1915+105 in the radio  and X-ray band
showed that the overall radio emission is correlated  with its X-ray
properties (eg: Harmon et al. 1997).  The radio emission from the source 
can be broadly
classified into three categories; (i) the relativistic superluminal radio
jets of flux density $\sim$ 1 Jy and above at cm wavelengths, 
with decay time-scales of several days
(Fender et al. 1999), (ii) the baby jets of 20 $-$ 40 min durations
with flux density of 20 $-$ 200 mJy both in infrared (IR) and radio
(Pooley \& Fender 1997; Eikenberry et al. 1998) and (iii) the plateau
state with persistent radio emission of 20 $-$ 100 mJy for extended
durations (Muno et al. 2001).  In the case of superluminal jets, the
radio emission has steep spectra and are observed at large distances
(400 $-$ 5000 AU) from the accretion disk (Fender et al. 1999; Dhawan
et. al 2000). The radio emission at this distance is believed to be
decoupled from the accretion disk. In contrast, the other two classes of
radio emission has flat spectra and occur close to the accretion disk
(within a few tens of AU; Muno et al. 2001). 
Even though superluminal jets and baby jets
are differentiated by their spectra, decay time scales, and the distance
from the accretion disk at which the emission takes place; there is an evidence
for the ejection of significant amount of relativistic material even during
the baby jets (Fender et. al 1999). The radio emission is believed to be
due to synchrotron emission from relativistic electrons and the dominant
decay mechanism is the adiabatic expansion losses (Mirabel et. al 1998).
The frequency corresponding to the peak radio flux in the source spectrum
depends on the energy and spectral distribution of the electrons in the
emitting volume. In the absence of in-situ acceleration, the energy of
the electrons will decrease both by radiative and expansion losses and
therefore, the peak flux will exhibit temporal evolution by shifting 
to lower
frequencies. In the case of emission of baby jets from the GRS1915+105,
where there is evidence for ejection of significant amount of relativistic
material, we expect low frequency ($<$ 1 GHz) radio
emission from the "relic" electrons (Kaiser et al. 2004).

In this paper, we present the low frequency radio data taken
simultaneously at two wavelengths and discuss the results in terms of
the geometric model of the source.

\section{Observations and Results}

The radio observations of GRS1915+105 were carried out at 610 and 244
MHz simultaneously 
with Giant Meterwave Radio Telescope (GMRT, Swarup et. al 1991).
These observations are part of major programme to monitor microquasars
with GMRT for extended duration (Pandey et al. 2004). The present data 
corresponds to two days of observations, viz, of June 21 and 22, 2003. 

GMRT is the world's largest radio telescope at meter-wavelengths  and
consists of 30 antennas, each of 45 metre diameter spread over about 25
km in the form of a compact central square and distributed Y-array. Some
of the  parameters of GMRT are given in Table 1, more details about
the telescope can be found in {\tt www.ncra.tifr.res.in}. GMRT has a
built-in facility to observe simultaneously at 610 and 244 MHz, which
was availed for the present observations. The flux density scale is set
by observing the primary
calibrator 3C286 or 3C48. A phase calibrator was observed before and after
a 30 min scan on GRS1915+105 for phase calibration. The integration time
was 16 s.  The data recorded from GMRT was converted to FITS and
was analysed using Astronomical Image Processing System ({\tt AIPS}). An
iterative method of phase self calibration was  performed to reduce
the phase errors and to improve the image quality.

\begin{table}
\begin{center}
\caption{Some useful parameters of GMRT}
\begin{tabular}{l l l l l l }
          &          &           &    &             &            \\
Frequency (MHz) & 151  & 235  & 325  & 610  & 1000 $-$ 1450      \\
Primary Beam ($^\circ$)& 3.8  & 2.5  & 1.8  & 0.9  & 0.56 - 0.4 \\
Resolution ($^{''}$)   &  20  & 13  &  9  &  5  & 2 - 3 \\
RMS (mJy)/hour  &   ??  & 2   & 1   & 0.5   &  0.1  \\
\end{tabular}
\end{center}
\end{table}

The radio images of GRS1915+105 at 244 and 610 MHz are presented in Figure 1.
The source is unresolved at both frequencies.
In  Figure 2, we have plotted the observed flux at different frequencies
during the two observations. The present observations clearly demonstrate
that GRS1915+105 exhibits strong radio emission at 610 and 244 MHz
and there is no evidence for spectral turnover up to 244 MHz. The flux
densities at these frequencies have been corrected for the increased
system temperature in the direction of GRS1915+105.  On June 21, 2003,
the flux density of GRS1915+105 at  610 MHz is 194.5$\pm$1.6  mJy and
at 244 MHz is
756$\pm$58.7 mJy giving a spectral index of ~$-$1.46 (S$_\nu \propto
\nu^\alpha$; Figure 2). On June 22, 2003, the flux density at 610 MHz
is 156.2$\pm$1.5 mJy and that at 244 is 672$\pm$7.9 mJy, with a spectral
index of ~$-$1.57 (Figure 2), marginally steeper than the previous day.
This marginal steepening is consistent with the expectation that the 244
MHz emission decays slower than 610 MHz.  The observed spectral indices
on both days suggests that the radio emission at low frequencies arises
in a optically thin medium.

\begin{figure}
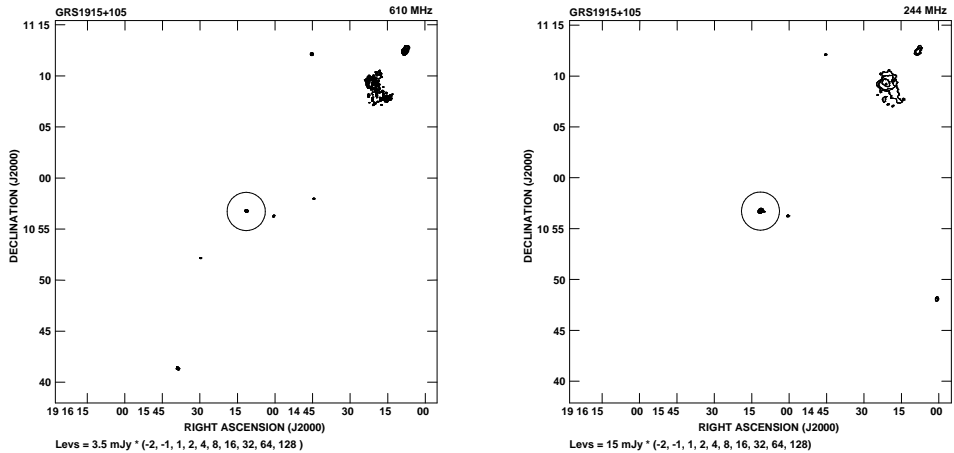

\hbox{
\hspace{0.15in}
\psfig{figure=fig1a.ps,height=2.5in}
\hspace{0.25in}
\psfig{figure=fig1b.ps,height=2.5in}
}
\caption{Field of GRS1915+105 at 610 (left) and 244 MHz (right).
The circle drawn at the central region is only to help to 
locate the GRS1915+105, which is at the center of this circle.
The rest of the sources in the field are likely to be unrelated
to GRS1915+105.}

\end{figure}

\section{Discussion}
The low frequency radio emission of microquasars are useful to understand
and constrain several parameters such as compactness of the radio emitting
plasma, the density of ambient medium.
The radiative lifetimes of the electrons are also
longer at these frequencies, thus the source will be visible for longer
duration as compared to higher radio frequencies. 

In the case of a compact radio emitting plasma, synchrotron self
absorption will result in the spectral turn around at lower
radio frequencies. In contrast, the extended diffused emission will result 
steep power-law spectra corresponding to a  optically thin region.
As seen in Figure 2,  in the case of GRS1915+105, the
flux density seems to increase at lower frequencies suggesting a large
population of low energy electrons in an optically thin plasma. The data
suggest that source is not self-absorbed and the large flux value does
indicate a separate region of large surface brightness in the source
geometry.
  
In the following, we discuss the possible geometry of the emission
and compare the radio properties of GRS1915+105 with other well known
microquasars.

\begin{figure}[t]
\hspace{0.35in}
\psfig{figure=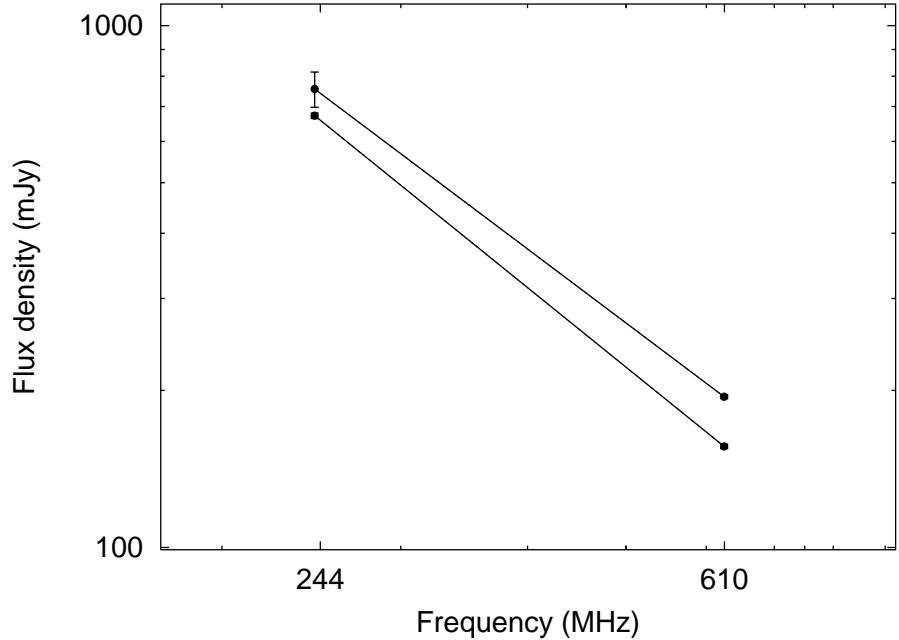,height=3.5in,angle=270}
\caption{Spectral index plot between 610 and 244 MHz for June 21 (upper
line) and June 22, 2003 (lower line)}
\end{figure}

\subsection{Proposed geometry of emission region}

The low frequency radio emission is unlikely to come from jet, since
the geometry of a jet is compact and therefore, the low frequency
radio emission would be self absorbed. In the standard model for the
spherical radio emitting plasma, the higher frequency emission comes
from the inner regions and the lower frequency comes from the outer
layers due to increased optical depth for the inner layers (van der Laan,
1966; Hjellming and Johnston, 1988).  Such a model is also referred as
"onion model".

In the case of GRS1915+105, the basic physical process will be similar to
the onion model with a difference that geometrical nature of the emitting
volume is not spherical, but a conical jet (Hjellming and Johnston,
1988). In this model, the infra-red emission occurs shortly after the
ejection of the plasma from the accretion disk and radio emission follows
with a time delay (Mirabel et al. 1998).  As the plasma moves outwards
along the jet, it also undergoes expansion and becomes optically thin.
The peak of the emission moves progressively to lower frequencies.
The plasma will continue to expand if the ambient medium is not dense
enough to slow down or stop the expansion.  Under such conditions, lobes
will be formed at the end of the jet, similar to case of AGNs (Figure 3).
Therefore no absorption effects will be visible and the spectra will
continue to be optically thin.

\subsection{Comparison with other microquasars}

Eventhough GRS1915+105 has been extensively observed over years in the
radio wavelengths, nearly all the observations have been  carried out
at higher radio frequencies.  In the absence of a spectral information
of the source at lower frequencies, the true nature of the emission
behavior is difficult to unravel, since the peak of the emission will
evolve in to lower frequency domain and the knowledge of the spectral
break is important. Some of the other known microquasars with strong
radio emission exhibit spectral turn over at cm wavelengths (Cyg X-3:
Miller-Jones et al. 2004; V4641 Sgr: Ishwara-Chandra and Pramesh Rao,
this proceedings).

In the case of Cygnus X-3, the radio spectra is optically thin and
steep above GHz frequencies whereas the spectra is flat or inverted at
lower frequencies. The spectral turnover is understood as synchrotron
self-absorption which suggests that the emitting region is compact.
This is possible if the expansion of the jet is limited by the dense
surrounding. Such clear spectral turn over at low radio frequencies
give important information about the source and its ambiance. If the
spectral turn over frequency is known, by applying the synchrotron self
absorption models, it is possible to estimate the size and the magnetic
field of the emitting region.  Similarly, the microquasar V4641 Sgr also
exhibits spectral turnover at low radio frequencies (Ishwara-Chandra and
Pramesh Rao, this proceedings).  However in the case of GRS1915+105,
the spectral turnover was not observed suggesting that the emitting
region must be extended.  In the present observations, GRS1915+105 is
unresolved at 610 MHz and at 244 MHz.  The low frequency radio images
are highly scatter broadened and scaling relation between the source size
and frequency need to be used to obtain the true size. High resolution,
high frequency imaging are required to obtain the scaling relation and
to estimate the true size of emitting plasma at low radio frequencies.

\begin{figure}[t]
\hbox{
\hspace{0.5in}
\vbox{
\vspace{-2cm}
\psfig{figure=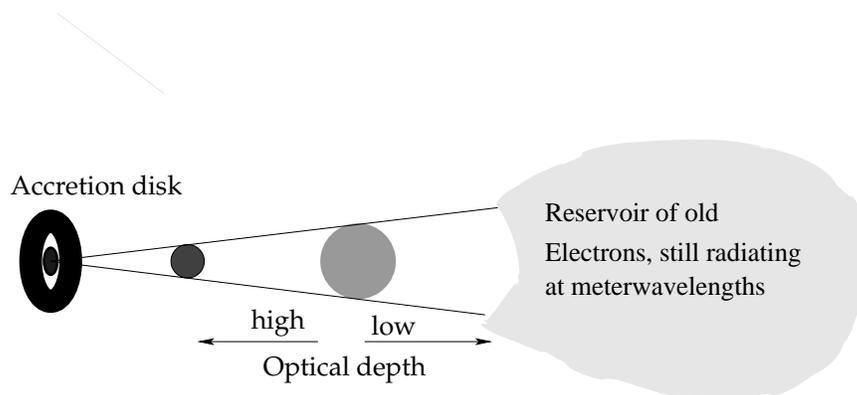,height=3.5in,angle=270}
}
}
\caption{Toy model to suggest the possible location for the 
low frequency radio emission.}
\end{figure}

\subsection{Is this unique to GRS1915+105?}

The related question to be addressed is whether the extended low
frequency radio emission in GRS1915+105 is present at times when there
are no flares at higher frequencies. Even in the case of baby jets and
plateau state where high frequency radio emission is coupled to the
activities of the accretion disk, the plasmon will expand outwards and
will eventually decouple. 
These plasmons are expected to radiate at lower radio
frequencies with longer lifetimes by virtue of the low energy of the
electrons and slower expansion rate. Hence it is important to monitor
GRS1915+105 at 610 and 244 MHz even in the radio quite state. As a part
of our ongoing project, GRS1915+105 has been observed at 610 and 244 MHz
even when there are no flares at high frequencies.  The data analysis
is in progress and the results will be published elsewhere.

\section{Summary}

We have presented for the first time, detailed low frequency radio
observations of the galactic microquasar GRS1915+105 with GMRT. The
observations have been carried out at 610 and 244 MHz simultaneously. 
Our results suggest that there is no spectral turn
over even at low radio frequencies of 244 MHz. We suggest that while the
radio emission at high radio frequencies could predominantly come from
compact jets, the lower frequency emission is likely to come from regions
far from the accretion disk. The relativistic plasma which was ejected
from the accretion disk will move outwards and in this process, it will
loose energy both by radiation and by expansion and will form the lobe,
similar to the case of AGNs.  This reservoir of low energy electrons is
likely to be the origin for optically thin low frequency radio emission.

\section*{Acknowledgments}

GMRT is run by the National Centre for Radio Astrophysics of the Tata
Institute of Fundamental Research. This research has made use of NASA's
Astrophysics Data System and of the SIMBAD database, operated at CDS,
Strasbourg, France.

\end{document}